\documentclass[11pt]{article} 

\usepackage[margin=1in]{geometry} 
\usepackage{setspace}            
\usepackage{amsmath, amssymb}    
\usepackage{graphicx}            
\usepackage{hyperref}            
\usepackage{authblk}             
\usepackage{natbib}

\usepackage{fancyhdr}
\begin{document}





\title{\textbf{ProtScan: Modeling and Prediction of RNA-Protein Interactions}}

\author[1]{Gianluca Corrado}
\author[2]{Michael Uhl}
\author[2,3]{Rolf Backofen}
\author[1]{Andrea Passerini}
\author[2,4]{Fabrizio Costa\thanks{To whom correspondence should be addressed: \texttt{f.costa@exeter.ac.uk}}}

\affil[1]{Department of Information Engineering and Computer Science, University of Trento, Trento, 38123, Italy}
\affil[2]{Department of Computer Science, University of Freiburg, Freiburg, 79110, Germany}
\affil[3]{BIOSS Centre for Biological Signalling Studies, Cluster of Excellence, University of Freiburg, Freiburg, 79110, Germany}
\affil[4]{Department of Computer Science, University of Exeter, Exeter EX4 4QF, UK}

\date{} 

\maketitle

\abstract{ \textbf{Motivation:} CLIP-seq methods are valuable techniques to
experimentally determine transcriptome-wide binding sites of RNA-binding
proteins. Despite the constant improvement of such techniques (e.g. eCLIP), the
results are affected by various types of noise and  depend on experimental
conditions such as cell line, tissue, gene expression levels, stress conditions
etc., paving the way for the {\em in silico} modeling of RNA-protein
interactions.\\ \textbf{Results:} Here we present ProtScan, a predictive tool
based on consensus kernelized SGD regression. ProtScan denoises and generalizes
the information contained in CLIP-seq experiments. It outperforms competitor
state-of-the-art methods and can be used to model RNA-protein interactions on a
transcriptome-wide scale.\\ \textbf{Availability:} ProtScan is available at
\href{https://github.com/gianlucacorrado/ProtScan/}{https://github.com/gianlucacorrado/ProtScan/}\\
\textbf{Contact:} \href{f.costa@exeter.ac.uk}{f.costa@exeter.ac.uk}\\ 


\section{Introduction}

RNA-binding proteins (RBPs) are regulating many vital aspects of transcribed
RNA such as splicing, maturation, stability and translation. Recent
studies~\citep{baltz2012mrna,castello2012insights,ray2013compendium,
gerstberger2014census} estimated that the human genome encodes more than 1500
RBPs. This implies that the complexity of post-transcriptional regulation by
RBPs is matching that of the regulation by transcription
factors~\citep{vaquerizas2009census}. An important step to understand this
layer of regulation is to detect binding sites of RBPs on regulated RNAs. In
this context, CLIP-seq (cross-linking and immunoprecipitation followed by next
generation sequencing)~\citep{licatalosi2008hits} together with its
modifications PAR-CLIP~\citep{hafner2010transcriptome},
iCLIP~\citep{konig2010iclip}, and eCLIP~\citep{van2016robust} has become the
state-of-the-art procedure for determining transcriptome-wide binding sites of
RBPs, resulting in libraries of reads that are bound and protected by a
specific RBP.

There are two important tasks in the {\em in silico} analysis of CLIP-seq data
for an individual RBP. The first obvious task is the reduction of the false
positive rate by removing reads that correspond to unspecific binding. This is
the task of peak calling, which determines regions that are enriched in
CLIP-seq reads and thus are likely bound by the RBP used in the CLIP-seq
experiment. Some popular peak calling tools are
PARalyzer~\citep{corcoran2011paralyzer},  Piranha~\citep{uren2012site} and
CLIPper~\citep{lovci2013rbfox}.

The second task is less obvious but possibly even more important, namely the
reduction of the false negative rate. False negatives are RNAs that are bound
by the RBPs but not detected in the CLIP-seq experiment. It turns out that the
false negatives are the major obstacle in using publicly available CLIP-seq
data to investigate biological questions related to RBPs. The reason is that
target sites identified by CLIP-seq depend on the transcriptome state of the
cells used for the CLIP-seq experiment, as the associated target transcripts
have to be expressed at certain levels to be detected. Thus, using these sites
for analyzing biological questions in cells with different transcriptome states
is suboptimal and can fail to detect many important sites, leading to wrong
biological conclusions. To give an example, \cite{ferrarese2014lineage}
investigated the role of the splice factor PTBP1 in differential splicing of
the tumor suppressor gene ANXA7 in glioblastoma. Albeit there was strong
biological evidence for PTBP1 directly binding ANXA7, no binding site was found
in a publicly available CLIP-seq dataset for PTBP1. Instead, only a
computational analysis was capable to detect and correctly localize the
presence of potential binding sites which were then experimentally validated.

In the last years, several specialized CLIP-seq databases have appeared like
doRiNA~\citep{anders2011dorina} and others ranging from those focused on in-
vitro and in-vivo experiments \citep{cook2010rbpdb} to those that offer
transcriptome-wide visualizations \citep{khorshid2010clipz}.
In the literature, approaches to predict RBP binding sites range from the
adaptation of tools built for DNA-binding motifs of transcription factors (see
\citep{das2007survey} for a survey) to specialized approaches that can take
the RNA folding properties into account. Well-known motif finder algorithms
of the first type include MEME~\citep{bailey1994fitting},
MatrixREDUCE~\citep{foat2006statistical} and
DRIMust~\citep{leibovich2013drimust}. Applications of these tools to the
analysis of RBPs are described in~\citep{sanford2009splicing, kazan2010rnacontext,
gupta2013hnrnp}.   Approaches developed explicitly for RNA motives include
MEMERIS~\citep{hiller2006using} (an extension of MEME biased towards single-
stranded RNA regions) and RNAContext~\citep{kazan2013rbpmotif} (which
distinguishes the type of unpaired regions between e.g. bulges, multiloops and
hairpins). mCarts~\citep{zhang2013prediction} is a tool based instead on a
Hidden Markov Model (HMM) that takes into account the number and the distance
between targets in addition to evolutionary conservation information (Zhang et
al. 2013). More recently
\cite{maticzka2014graphprot} introduced a graph kernel approach showing
improved performance over motif-based techniques.

Here we propose a novel method called ProtScan that uses a combination of
kernelized regression with consensus voting in order to localize the protein
binding sites. The key idea is to cast the identification of target regions in
long RNA sequences as a regression task over short {\em moving} windows, where
the regressed information is the distance to the closest target. It is well
known~\citep{hansen1990neural,
perrone1993improving,krogh1995neural,breiman2001random} that when the
classifiers in an ensemble are both individually strong and collectively {\em
diverse}, the consensus prediction is on average better than that of any
individual classifier. In our case we want
to train different predictors with sequences that are at different distances
from the target site. The intuition is that at different distances there will
be different clues that could hint at the presence of a nearby target site.
Finally, instead of having multiple classifiers, one for each different
distance value, we train a single regressor to directly predict the distance
value.


After introducing the details of the proposed method in
Section~\ref{sec:method}, in Section~\ref{sec:res} we empirically investigated
the following questions: Q1) is there an advantage in formulating the problem  as a regression task rather than a standard classification task? Q2) is the feature representation obtained by string
kernel methods competitive w.r.t. other state-of-the-art approaches (e.g.
based on deep artificial neural networks)?


\section{Materials and Methods}
\label{sec:method}

The ProtScan pipeline is composed of several steps that can be aggregated into
two main components: the first one models RNA-protein interactions and is used to
predict the interaction profiles, while the second identifies binding sites as
significant peaks in these profiles. In the following we give a brief overview of
the processing flow (see Figure~\ref{fig:train_test}).

The first component estimates RNA-protein interaction profiles using a
combination of kernelized regression with consensus voting.
The kernelized regression has the task to estimate the distance of a
portion of the RNA from the closest binding site, while the consensus voting is
used to aggregate the predictions from the different regions.
\begin{figure*}[!tpb]
 \centering
 \includegraphics[width=\textwidth]{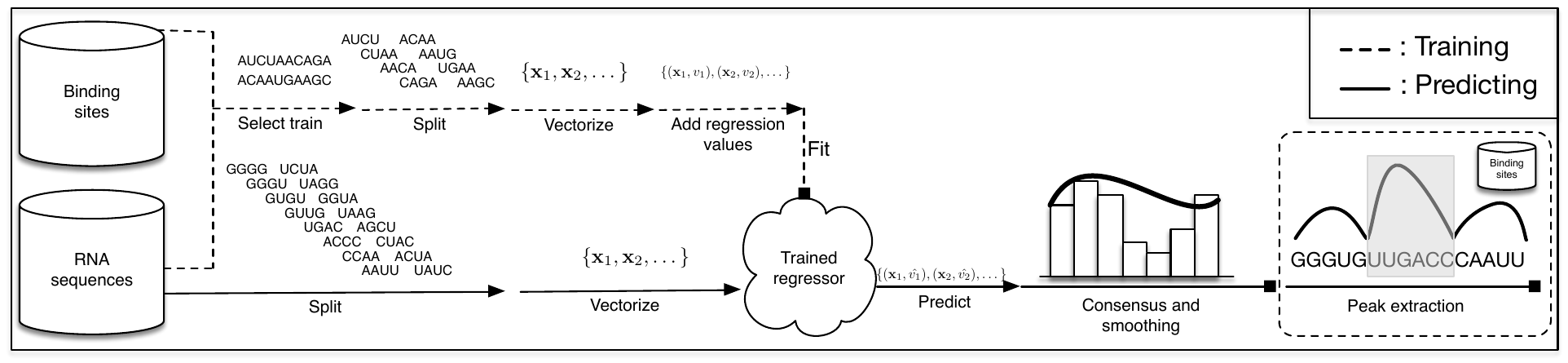}
 \caption{Workflow depicting the steps required for training a ProtScan model
 (dashed lines) and for predicting interaction profiles using a trained model
 (solid lines). The dashed box represents the peak extraction step.}
 \label{fig:train_test}
\end{figure*}
The regressor is trained using experimentally verified binding sites (dashed
lines in Figure~\ref{fig:train_test}). First,  we select an informative set of
fixed-length RNA fragments for training. We distinguish positive fragments, when
these are centered on a protein binding site, and negative fragments, when these
are sampled  at random in RNA regions that are far from binding sites. The
fragments are further split into smaller overlapping windows, which are
transformed into sparse vectors using a kernelized approach. We annotate each
window with its distance from the closest binding site, and use a special value
for the negative windows.
Finally, we learn a regressor to
predict the association between windows and their distance to the closest
binding site.

In the test phase, we compute the interaction profile for a set of arbitrarily
long RNA sequences (solid lines in Figure~\ref{fig:train_test}). First, each
RNA is split into small overlapping windows. The windows are then mapped to
vectors, and their distance from the closest
binding site is assessed using the trained regressor. All distances are then
aggregated in a histogram for consensus voting. Finally, the counts are smoothed
to obtain the RNA-protein interaction profile with single-nucleotide resolution.

The second component extracts the most reliable interactions from the
predicted profiles. It can therefore be used to {\em denoise} a CLIP-seq
experiment, removing protocol artifacts and biases. Starting from the predicted
profiles generated by the first component, ProtScan identifies all the peaks and
then, using the experimental evidence as control, selects peaks
above a desired significance level (dashed box in Figure~\ref{fig:train_test}).

\subsection{Dataset}
\label{sec:dataset}

We use data obtained with the enhanced CLIP (eCLIP)~\citep{van2016robust}
protocol. We downloaded BED narrowPeak files, containing the output of the
analysis pipeline of human eCLIP experiments, from the ENCODE project
website~\citep{sloan2016encode} (April 2016 release). The BED narrowPeak files
contain the genomic coordinates of RBP binding regions and their respective {\em
fold change} values, i.e. the base 2 logarithm of the ratio between the number
of aligned reads in the CLIP and the ones in the RNAseq control library.
Higher fold change values are indicative of more reliable binding regions.

The full dataset includes 96 RBPs, with experiments performed in two different
cell lines, i.e. K562 and HepG2 (38 RBPs on both cell lines, 40 only on K562
and 18 only on HepG2). Each experiment, identified by a protein and a cell
line, is performed in two replicates. The presence of two replicates allows
us to perform quality control on the data and it allows us to select only stable
experiments. We define {\em binding sites} as regions with a fold change higher than
a user-defined threshold. By setting the threshold to 2.0 and 3.0 respectively, we
identify two increasingly stringent sets of binding sites. 
This allows us to compare the robustness of predictive methods
for different levels of data quality.
For each set, we then discard experiments where the total number of binding
sites across the two replicates varies by more than 15\%. A subset of 46
different RBPs passes this quality control, 8 having experiments on both cell
lines, 25 only on K562 and 13 only on HepG2. When looking at the fold change
threshold, 20 RBPs pass the quality control at both values, 22 only at 2.0 and
4 only at 3.0.

The BED narrowPeak files report the binding regions in genomic coordinates
(hg19 assembly), but for the scope of this work we are interested in working
with full-length gene sequences. First, genomic coordinates are converted from
hg19 to hg38 assembly using the UCSC's liftOver tool~\citep{speir2016ucsc}.
Afterwards, genomic coordinates are converted to gene coordinates using the
human cDNA GTF file from Ensembl as a reference (release
84)~\citep{yates2015ensembl}.

More formally, for each RNA sequence $r$ we define the set $\mathcal{B}^r$ of
coordinates $b$, where $b = (e-s)/2$ is the center of a binding site on $r$
that starts at coordinate $s$ and ends at coordinate $e$. If an RNA sequence
$r$ has no binding sites, then $\mathcal{B}^r = \varnothing$.

\subsection{Interaction profile predictor}
\label{sec:interaction_profile}

In the following we detail the steps for the
RNA-protein interaction profile estimator (Figure~\ref{fig:train_test}).

\paragraph{Selection of training subsequences.}

Training subsequences are selected in order to include information surrounding
experimentally determined binding sites (positive RNA subsequences) as well as
``background'' information from RNA portions far away from any binding site
(negative RNA subsequences). Each positive subsequence is centered on a
binding site and is extended $d_{max}$ nucleotides on both sides for a total
length of $2 d_{max}$. Negative subsequences have the same length but are
centered on nucleotides more than $d_{max}$ nucleotides away from the center of
any binding site. Since including a huge number of negatives that overwhelms
the number of positives might cause improper training of the regressor, we
select at random a number of negative subsequences that is proportional to the
number of positive ones ($negative\_ratio$ times the number of positives). Both
$d_{max}$ and $negative\_ratio$ are hyperparameters of the model that can be
optimized with the the random search hyperparameter optimization procedure of
ProtScan (see Supplementary material).

\paragraph{Splitting.}

Each sequence $r$ of length $l$ (when considering training subsequences
$l = 2 d_{max}$) is split in overlapping windows of size $split\_window <
l$. Each window is identified by the position $i$ of its central nucleotide on
$r$. The amount of overlap between two consecutive windows is controlled by the
parameter $split\_step$ with $split\_step < split\_window$ (the strict inequality ensures overlap).

\paragraph{Vectorization.}

A typical approach to process non-vector data (such as sequences or graphs) is
to employ the "kernel trick"~\citep{shawe2004kernel}. The trick consists in
using an algorithm that interacts with the input only in terms of inner
product between instances. All that is needed then is a way to efficiently
define an inner product between discrete sequences. A typical solution is
offered by string kernels~\citep{leslie2002spectrum} that compute the fraction
of common $k$-mers (i.e. short subsequences of length $k$). Here, for
efficiency reasons, we use a different approach and compute an explicit
feature mapping from discrete sequences $x$ to sparse vectors in very high
dimensional spaces $\mathbb{R}^d$, where $d$ is typically in the order of tens
of thousands. To do this we follow the Neighborhood Subgraph Pairwise Distance
Kernel (NSPDK) approach proposed in~\citep{costa2010fast} and define a feature
construction procedure $\phi_k(x) \mapsto \mathbb{R}^d$ that returns the
histogram of the occurrences of each $k$-mer in a string $x$. A hash function
$h: \Sigma^* \mapsto \mathbb{N}$ is used  to map $k$-mers (short strings in a
finite alphabet $\Sigma$) to  the corresponding integer codes $n \in
\mathbb{N}$ in the addressable space (i.e. $n< d$). In order to take into
account the contribution of $k$-mers of different complexities (different
values of $k$) in a balanced way, we first consider the normalized version:
$\hat{\phi}_k(x) = \phi_k(x)/\sqrt{<\phi_k(x) \phi_k(x)>}$, then we combine
the vector representations for different orders $k$ in a single vector:
$\phi_C(x)=\sum_{k=0}^C \hat{\phi}_k(x)$ and finally we output the normalized
result: $\hat{\phi}_C(x) = \phi_C(x)/\sqrt{<\phi_C(x) \phi_C(x)>}$.  We call
the maximum $k$-mer size $C$, the {\em complexity} of the vectorization.  When
$C$ is small (say below 10), the number of non-zero features per sequence is
small (there can be maximally as many distinct features as $C$ times the
length of the sequence), yielding an efficient computation for the downstream
learning algorithm. Note that the normalization changes the representation
from a list of counts to a list of proportions. In practical terms this means
that instead of considering the number of occurrences of a k-mer in a sequence
we consider its frequency. To fix ideas, when $C=6$ (the value which was found
optimal by the cross validation selection procedure on most RBP tasks in our
experimental evaluation), NSPDK will generate features for $k \in
\{1,2,3,4,5,6\}$, i.e. ranging from the nucleotide composition ($k=1$), to the
di-nucleotide composition, up to the 6-mers composition. Note that larger
values of $k$ increase the complexity of the method until it starts to
memorize idiosyncratic aspects of the training data and the predictive
performance on future instances degrades.

\begin{figure}
 \centering
 \includegraphics[width=0.95\columnwidth]{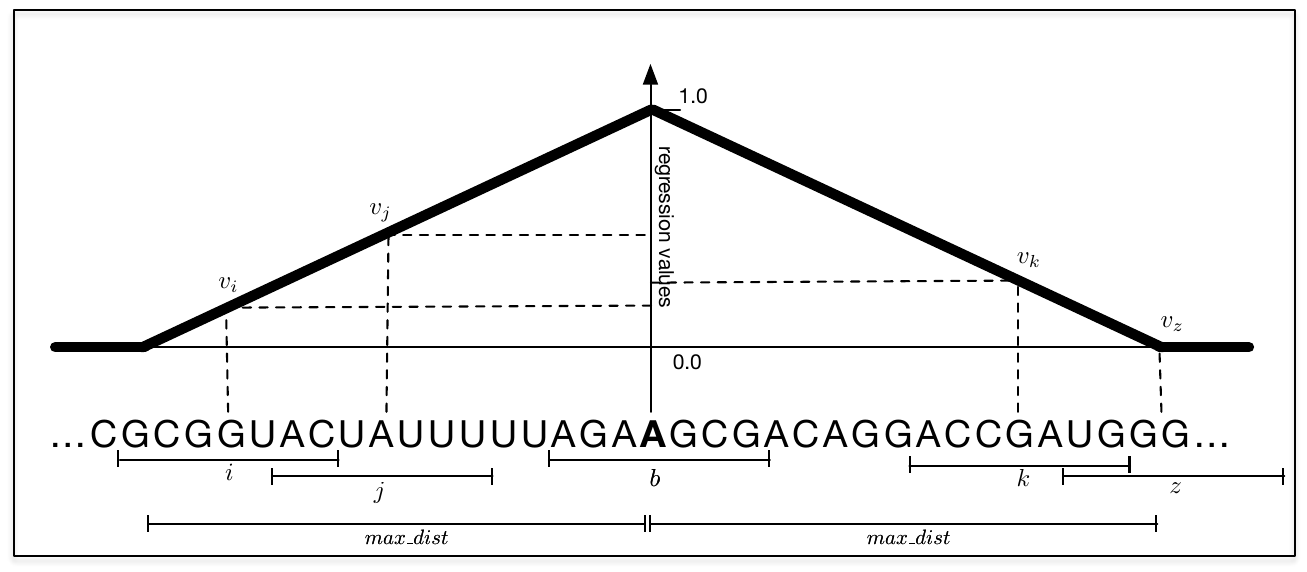}
 \caption{Example of the definition of regression values. The regression value
 $v_i$ is lower than $v_j$ because window $i$ is
 farther from the target site $b$ than window $j$. Although window $k$ is
 positioned upstream from the binding site $b$ and windows $i$ and $j$ are
 downstream, $v_i < v_k < v_j$ because the regression values do not need
 to account for the relative position w.r.t. the binding site but only for
 the absolute distance. Moreover, $v_z = 0$ because the center of the window $z$
 is located at $d_{max}$ nucleotides from the binding site $b$.}
 \label{fig:regression_vals}
\end{figure}

\paragraph{Regression.}
In ProtScan we employ a ridge regressor with squared loss and $l_2$
regularization trained using stochastic gradient descent (SGD). Let $i$ be the
center of a window of a RNA sequence $r$, then $v_{i}$ is the corresponding
regression value which is inversely proportional to the distance of $i$ from the
closest binding site on sequence $r$, if $i$ is a positive window, and zero
otherwise

\begin{equation}
    \label{eq:dist_to_val}
    v_i = \bigg\{
    \begin{array}{@{}ll@{}}
        \textrm{max}(0, 1 - \frac{\min_{b \in \mathcal{B}^r}|i - b|}{d_{max}}) & \text{if $\mathcal{B}^r \neq \emptyset$}\\
        0 & \text{otherwise}
    \end{array}
 \end{equation}

In the prediction step, we estimate the distance values for RNA windows of test
RNA sequences. For each test window $i$, the regressor predicts a value
$\hat{v}_i$. The predicted value is mapped to a
distance $\hat{d}_i \in [0, d_{max}]$ inverting
Equation~\ref{eq:dist_to_val}:

\begin{equation}
    \label{eq:val_to_dist}
    \hat{d}_i = d_{max} * (1 - \hat{v}_i)
\end{equation}

Note that Equation~\ref{eq:dist_to_val} assigns regression
values according to the {\em absolute} value of the distance from the most
adjacent binding site (Figure~\ref{fig:regression_vals}) and that it cannot recover
the relative position of the window with respect to the binding site (i.e.
downstream or upstream). Encoding directionality information using, for example,
negative regression values to indicate upstream locations yielded poor
performance due to the discontinuity at zero. As we will show now, the exact
location can be recovered using a consensus voting procedure.

\paragraph{Consensus voting and smoothing.}

In the test phase we aggregate the predictions from all available windows.  We
build a histogram $\mathbf{h} = (h_1, \dots, h_l)$, where $l$ is the length of a
test RNA sequence $r$ and $h_j$ aggregates the votes received by its $j$-th
nucleotide. We discard a window $i$ if $\hat{v}_i \leq 0$ as it is predicted to
be too far from a binding site to be relevant. Otherwise, every prediction
contributes two votes, one upstream to position $i-\hat{d}_i$ and one downstream
to $i+\hat{d}_i$ (recall that the regressor is trained over the absolute value
of the distance). Votes for position $j$ are thus computed as:

\begin{equation}
    \label{eq:consensus}
    h_j = \sum_{i \in windows(r)} \Bigg\{
    \begin{array}{@{}ll@{}}
        \hat{v}_i & \text{if $i \pm \hat{d}_i = j$ and $\hat{d}_i < d_{max}$}\\
        0 & \text{otherwise}
    \end{array}
    \qquad \forall j
\end{equation}

In Equation~\ref{eq:consensus} each vote is weighted according to the
predicted distance, i.e. the closer the voting window the higher the weight.
This is done to impose a bias whereby RNA windows that are closer to a binding
site are considered more important for the protein recognition than more
distant windows. Development experiments (not reported) indicate that vote
weighting has a marginal but beneficial effect on the predictive performance
estimate. Secondly, the vote is added to both $i \pm \hat{d}_i$, i.e.
upstream and downstream from the window coordinate. At first glance, this seems
an issue, as one of the two votes is clearly wrong. However, votes will combine
in a constructive way only on the true location while they will incoherently
spread out in the other direction (see Figure~\ref{fig:voting}).

\begin{figure}[!tpb]
 \centering
 \includegraphics[width=\columnwidth]{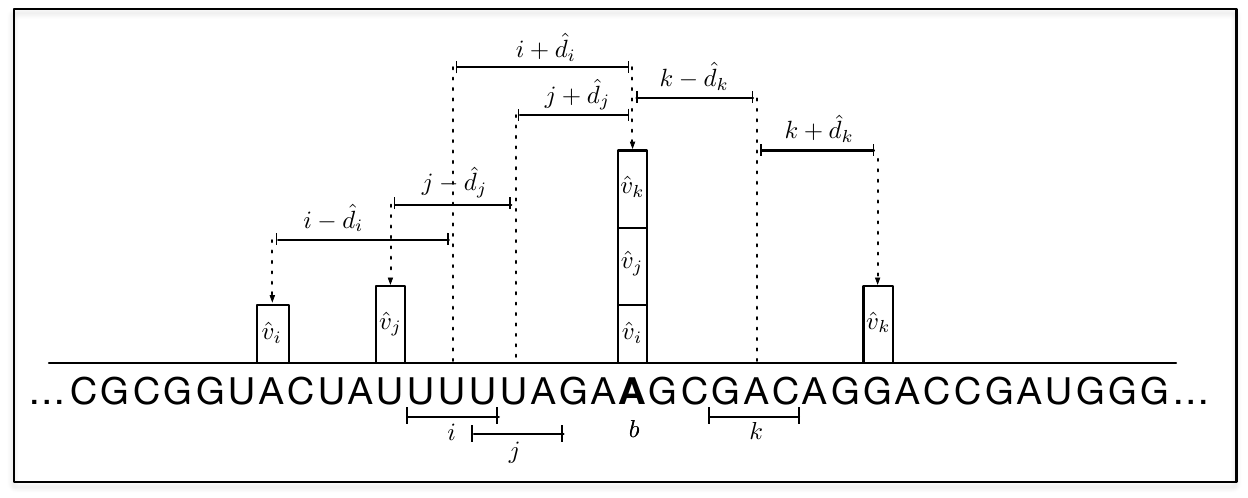}
 \caption{Example of the consensus voting approach. Under the assumption that
 the regressor is perfectly trained ($\hat{v}_c = v_c$ for all windows $c$),
 $\hat{d}_c$ represents the exact distance of a window $c$ to the closest binding site $b$. From the example it is possible to notice that the votes are
 correctly piling up on the binding site and spreading on the other
 nucleotides.}
 \label{fig:voting}
\end{figure}

Finally, Gaussian smoothing, i.e. the convolution of histogram $\mathbf{h}$ with
a Gaussian  $\mathcal{N}(\mu, \sigma)$, is applied to the histogram $\mathbf{h}$
to denoise it and to produce a single-nucleotide resolution interpolated profile.

\paragraph*{Hyperparameter selection.}

ProtScan exhibits a relatively large set of hyperparameters (see Supplementary
Material), which we optimize using a two-fold cross validation random search
approach~\citep{bergstra2012random}. By running the hyperparameter optimization
over 34 models for 11 different RBPs for the comparison with the related work
(Section~\ref{ssec:comparison}), we noted that several optimal
hyperparameter values where stable for a wide range of RBPs (see Supplementary
Material for details). These results are now incorporated as default parameters
and allow to train ProtScan on novel RBPs, skipping the computationally expensive
hyperparameters optimization phase while maintaining high predictive
performance.

\subsection{Peak extraction}
\label{sec:peak}

Predicted interaction profiles consist of single-nucleotide resolution signals
indicative of the RNA-protein coupling. However, the localization of significant
peaks in these profiles is a non-trivial task, akin the process of peak calling in CLIP-seq data analysis. We therefore implemented a similar approach to find significant peaks and thus sites likely bound by the RBP from the predicted interaction profiles.

To do so we first extract all the peaks in the predicted profiles using a
variant of the {\em mean shift} algorithm~\citep{comaniciu2002mean}. Mean shift scans
a sequence with a fixed-length sliding window and records the maximum value found
in each window. It then iteratively repeats the procedure over the sequence of
maxima found until no further change occurs. An analogous procedure is used to
localize all the minima. After identifying all the local maxima and minima in
the profile, we define a candidate predicted binding site as a {\em block} $b = (s,
e)$ with coordinates $(s, e) : s < e$. If both $s$ and $e$ are minima, the
block contains no other minimum and at least one maximum.

In order to select the subset of significant binding sites among the extracted
peaks, we compare them with a background distribution fit on negative data.
First, a cumulative Gaussian  distribution for the maximum is fit over the
height of the blocks coming from transcripts without experimental evidence of
binding (negative examples). Second, each candidate block is accepted as
significant if it stays in the top $\theta^{th}$ percentile of the distribution,
with $\theta$ specified by the user. The procedure is cross validated two-fold
to avoid overfitting.


\section{Results and discussion}
\label{sec:res}

Amongst popular methods to model target regions for RBPs there are approaches
based on Position Specific Score Matrices (PSSM), such as
MEME~\citep{bailey1994fitting} or DREME~\citep{bailey2011dreme}. In the
following we exploited the results reported in \cite{maticzka2014graphprot} that
show how more sophisticated approaches that can model higher order
correlations between nucleotides exhibit greater predictive performance and hence
compare directly against GraphProt~\citep{maticzka2014graphprot}.

Unfortunately the comparison to other kernel approaches (such as the spectrum
kernel of~\cite{leslie2002spectrum}) is hindered by computational complexity
issues. Kernel methods that do not produce an explicit feature encoding, in
fact, cannot be easily applied to transcriptomic or genomic scale data since
their complexity grows quadratically with the number of fragments or regions
considered. For this reason we resorted to compare ProtScan with a
state-of-the-art method based on a deep neural network architecture called
DeepBind~\citep{alipanahi2015predicting}.

\subsection{Comparison with related work}
\label{ssec:comparison}

GraphProt~\citep{maticzka2014graphprot} is an SVM-based
classifier with a graph kernel developed for RNA
molecules, trained to predict sequence- and
structure-binding preferences of RNA-binding proteins from high-throughput
experimental data. DeepBind~\citep{alipanahi2015predicting} uses deep
convolutional neural networks (CNNs) to model RNA-protein binding patterns from
RNAcompete data~\citep{ray2009rapid}.

The comparison was performed on a subset of 11 RBPs from our dataset
(Section~\ref{sec:dataset}) for which a pre-trained DeepBind model is available
(training a DeepBind model is computationally very expensive and requires
powerful hardware like a GPU cluster). For each protein, multiple tests were
performed considering different cell lines, fold change values to define
binding sites from experimental evidence, and technical replicates, for a
total of 34 comparisons. In this comparison, we tested the performance of the
three approaches on $\sim$ 1\% of human genome ($\sim$ 600 protein coding and
non-coding genes). The test genes were selected at random, keeping the same
ratio between bound and unbound genes that is present in the full dataset.

\begin{table}
\caption{Performance comparison among GraphProt~\citep{maticzka2014graphprot},
DeepBind~\citep{alipanahi2015predicting} and our method, ProtScan, on
11 RBPs. For each test protein, multiple tests were performed taking into
consideration different cell lines (CL), fold changes (FC),
and replicates (R), for a total of 34 comparisons. For each comparison, the best score is highlighted in boldface.}
\begin{tabular}{|c|c|c|c|r|r|r|}
\hline
\multicolumn{ 4}{|c|}{\textbf{}} & \multicolumn{ 3}{c|}{\textbf{AUC ROC}} \\ \hline
\multicolumn{1}{|c|}{\textbf{RBP}} & \multicolumn{1}{c|}{\textbf{CL}} & \multicolumn{1}{c|}{\textbf{FC}} & \multicolumn{1}{c|}{\textbf{R}} & \multicolumn{1}{c|}{\textbf{GraphProt}} & \multicolumn{1}{c|}{\textbf{DeepBind}} & \multicolumn{1}{c|}{\textbf{ProtScan}} \\ \hline
\multicolumn{ 1}{|c|}{FMR1} & \multicolumn{ 1}{c|}{K562} & \multicolumn{ 1}{c|}{2.0} & 1 & 0.83 & 0.63 & \textbf{0.88} \\ \cline{ 4- 7}
\multicolumn{ 1}{|l|}{} & \multicolumn{ 1}{l|}{} & \multicolumn{ 1}{l|}{} & 2 & 0.80 & 0.62 & \textbf{0.84} \\ \hline
\multicolumn{ 1}{|c|}{GTF2F1} & \multicolumn{ 1}{c|}{HepG2} & \multicolumn{ 1}{c|}{2.0} & 1 & 0.71 & 0.56 & \textbf{0.80} \\ \cline{ 4- 7}
\multicolumn{ 1}{|l|}{} & \multicolumn{ 1}{l|}{} & \multicolumn{ 1}{l|}{} & 2 & 0.78 & 0.58 & \textbf{0.86} \\ \cline{ 3- 7}
\multicolumn{ 1}{|l|}{} & \multicolumn{ 1}{l|}{} & \multicolumn{ 1}{c|}{3.0} & 1 & 0.72 & 0.56 & \textbf{0.79} \\ \cline{ 4- 7}
\multicolumn{ 1}{|l|}{} & \multicolumn{ 1}{l|}{} & \multicolumn{ 1}{l|}{} & 2 & 0.80 & 0.58 & \textbf{0.86} \\ \hline
\multicolumn{ 1}{|c|}{HNRNPA1} & \multicolumn{ 1}{c|}{HepG2} & \multicolumn{ 1}{c|}{2.0} & 1 & 0.72 & 0.76 & \textbf{0.81} \\ \cline{ 4- 7}
\multicolumn{ 1}{|l|}{} & \multicolumn{ 1}{l|}{} & \multicolumn{ 1}{l|}{} & 2 & 0.72 & 0.75 & \textbf{0.82} \\ \cline{ 2- 7}
\multicolumn{ 1}{|l|}{} & \multicolumn{ 1}{c|}{K562} & \multicolumn{ 1}{c|}{2.0} & 1 & 0.72 & 0.77 & \textbf{0.80} \\ \cline{ 4- 7}
\multicolumn{ 1}{|l|}{} & \multicolumn{ 1}{l|}{} & \multicolumn{ 1}{l|}{} & 2 & 0.72 & 0.74 & \textbf{0.83} \\ \cline{ 3- 7}
\multicolumn{ 1}{|l|}{} & \multicolumn{ 1}{l|}{} & \multicolumn{ 1}{c|}{3.0} & 1 & 0.71 & \textbf{0.77} & \textbf{0.77} \\ \cline{ 4- 7}
\multicolumn{ 1}{|l|}{} & \multicolumn{ 1}{l|}{} & \multicolumn{ 1}{l|}{} & 2 & 0.72 & 0.74 & \textbf{0.80} \\ \hline
\multicolumn{ 1}{|c|}{HNRNPC} & \multicolumn{ 1}{c|}{HepG2} & \multicolumn{ 1}{c|}{2.0} & 1 & 0.68 & \textbf{0.86} & \textbf{0.86} \\ \cline{ 4- 7}
\multicolumn{ 1}{|l|}{} & \multicolumn{ 1}{l|}{} & \multicolumn{ 1}{l|}{} & 2 & 0.71 & 0.77 & \textbf{0.87} \\ \hline
\multicolumn{ 1}{|c|}{HNRNPK} & \multicolumn{ 1}{c|}{K562} & \multicolumn{ 1}{c|}{3.0} & 1 & 0.81 & 0.86 & \textbf{0.89} \\ \cline{ 4- 7}
\multicolumn{ 1}{|l|}{} & \multicolumn{ 1}{l|}{} & \multicolumn{ 1}{l|}{} & 2 & 0.79 & 0.85 & \textbf{0.90} \\ \hline
\multicolumn{ 1}{|c|}{IGF2BP2} & \multicolumn{ 1}{c|}{K562} & \multicolumn{ 1}{c|}{2.0} & 1 & 0.75 & 0.29 & \textbf{0.80} \\ \cline{ 4- 7}
\multicolumn{ 1}{|l|}{} & \multicolumn{ 1}{l|}{} & \multicolumn{ 1}{l|}{} & 2 & 0.76 & 0.31 & \textbf{0.79} \\ \hline
\multicolumn{ 1}{|c|}{IGF2BP3} & \multicolumn{ 1}{c|}{HepG2} & \multicolumn{ 1}{c|}{2.0} & 1 & 0.66 & 0.35 & \textbf{0.85} \\ \cline{ 4- 7}
\multicolumn{ 1}{|l|}{} & \multicolumn{ 1}{l|}{} & \multicolumn{ 1}{l|}{} & 2 & 0.70 & 0.35 & \textbf{0.82} \\ \hline
\multicolumn{ 1}{|c|}{KHDRBS1} & \multicolumn{ 1}{c|}{K562} & \multicolumn{ 1}{c|}{2.0} & 1 & 0.60 & 0.64 & \textbf{0.64} \\ \cline{ 4- 7}
\multicolumn{ 1}{|l|}{} & \multicolumn{ 1}{l|}{} & \multicolumn{ 1}{l|}{} & 2 & 0.62 & 0.63 & \textbf{0.66} \\ \hline
\multicolumn{ 1}{|c|}{QKI} & \multicolumn{ 1}{c|}{HepG2} & \multicolumn{ 1}{c|}{2.0} & 1 & 0.59 & 0.68 & \textbf{0.74} \\ \cline{ 4- 7}
\multicolumn{ 1}{|l|}{} & \multicolumn{ 1}{l|}{} & \multicolumn{ 1}{l|}{} & 2 & 0.54 & 0.56 & \textbf{0.74} \\ \cline{ 3- 7}
\multicolumn{ 1}{|l|}{} & \multicolumn{ 1}{l|}{} & \multicolumn{ 1}{c|}{3.0} & 1 & 0.58 & 0.68 & \textbf{0.72} \\ \cline{ 4- 7}
\multicolumn{ 1}{|l|}{} & \multicolumn{ 1}{l|}{} & \multicolumn{ 1}{l|}{} & 2 & 0.54 & 0.56 & \textbf{0.69} \\ \hline
\multicolumn{ 1}{|c|}{TARDBP} & \multicolumn{ 1}{c|}{K562} & \multicolumn{ 1}{c|}{2.0} & 1 & 0.71 & 0.84 & \textbf{0.88} \\ \cline{ 4- 7}
\multicolumn{ 1}{|l|}{} & \multicolumn{ 1}{l|}{} & \multicolumn{ 1}{l|}{} & 2 & 0.72 & 0.86 & \textbf{0.88} \\ \hline
\multicolumn{ 1}{|c|}{U2AF2} & \multicolumn{ 1}{c|}{HepG2} & \multicolumn{ 1}{c|}{2.0} & 1 & 0.59 & 0.68 & \textbf{0.76} \\ \cline{ 4- 7}
\multicolumn{ 1}{|l|}{} & \multicolumn{ 1}{l|}{} & \multicolumn{ 1}{l|}{} & 2 & 0.59 & 0.68 & \textbf{0.72} \\ \cline{ 2- 7}
\multicolumn{ 1}{|l|}{} & \multicolumn{ 1}{c|}{K562} & \multicolumn{ 1}{c|}{2.0} & 1 & 0.59 & 0.69 & \textbf{0.78} \\ \cline{ 4- 7}
\multicolumn{ 1}{|l|}{} & \multicolumn{ 1}{l|}{} & \multicolumn{ 1}{l|}{} & 2 & 0.62 & 0.67 & \textbf{0.79} \\ \cline{ 3- 7}
\multicolumn{ 1}{|l|}{} & \multicolumn{ 1}{l|}{} & \multicolumn{ 1}{c|}{3.0} & 1 & 0.59 & 0.69 & \textbf{0.76} \\ \cline{ 4- 7}
\multicolumn{ 1}{|l|}{} & \multicolumn{ 1}{l|}{} & \multicolumn{ 1}{l|}{} & 2 & 0.62 & 0.67 & \textbf{0.77} \\ \hline
 &  &  & \multicolumn{1}{l|}{\textbf{AVG}} & \textbf{0.69} & \textbf{0.65} & \textbf{0.80} \\ \hline
 &  &  & \multicolumn{1}{l|}{\textbf{STD}} & \textbf{0.08} & \textbf{0.15} & \textbf{0.07} \\ \hline
\end{tabular}
\label{tab:comparison}
\end{table}

We computed the interaction profiles for ProtScan as explained in
Section~\ref{sec:interaction_profile}. GraphProt and DeepBind were originally
designed for a binary classification task, i.e. to identify whether a RNA sequence
contains a RBP binding site, but they were not designed to model interactions at
single-nucleotide resolution. Despite that, GraphProt allows to partition the
feature set according to each nucleotide. This enables the marginalization of
the weights induced by the linear estimator and extract a nucleotide-wise
interaction profile. We adapted DeepBind using an approach similar to
the one used in ProtScan, i.e. we split the RNA in overlapping windows and we
aggregated the score of all the windows (see the Supplementary Material for
details).

Note that modeling binding sites at single-nucleotide resolution is a much
harder task than the binary interaction classification prediction. First, the
amount of predictions is significantly higher: from one
per RNA sequence to one per nucleotide. Second, and more importantly, when
considering long RNA sequences, the ratio between interacting and non-interacting
nucleotides is extremely small, resulting in a drastically unbalanced prediction
task (the average ratio in our dataset is 1 to 2500).

Table~\ref{tab:comparison} reports the AUC ROC results for each RBP, computed
over all target transcripts and positions. AUC ROC measures the quality of the
ranking, allowing to compare interaction profiles on different scales. In
fact, a perfect AUC ROC score of 1.0 is achievable only when all the
interacting nucleotides of the test sequences obtained higher interaction
scores than the non-interacting ones. It is important to note that the AUC ROC
values reported in~\cite{maticzka2014graphprot}
and~\cite{alipanahi2015predicting} are on a different task, namely on a
balanced binary classification. The experiments show that ProtScan outperforms
the other approaches nearly always (except for two cases in which it ties with
DeepBind). With an average AUC ROC of 0.8, ProtScan achieves a relative AUC
ROC improvement of 35\% over GraphProt, and of 43\% over DeepBind. Performing
a paired analysis, we observed that ProtScan outperforms GraphProt in all 34
cases, and DeepBind in 32 out of 34 ones (with two ties). The paired analysis
between GraphProt and DeepBind revealed that the former has a better average
AUC ROC while the latter wins in terms of pairwise comparisons (24 wins out of
34 cases). We note that the average AUC ROC of DeepBind was severely penalized
by the scores obtained for two proteins (IGF2BP2 and IGF2BP3). It has been
speculated that the IGF2BP family proteins (IGF2BP1-3) possess more complex
binding patterns due to the presence of multiple RNA- binding
domains~\citep{hafner2010transcriptome}. In fact it was shown for IGF2BP1 that
the RBP relies on two RNA-binding sites, which can be found at varying
distances and orientations in functional target
sequences~\citep{patel2012spatial}. However, since DeepBind models were
trained on RNAcompete data, they rely on local sequence motifs, which could be
insufficient to represent two or more disconnected binding sites. The good
predictive score of ProtScan indicates that the use of a much larger sequence 
context can help to identify these complex binding
patterns also in the case of IGF2BP family proteins.

In the supplementary material we also report the area under
the precision recall curve (auPRC), which is a performance measure suitable for highly
unbalanced predictive tasks. The results indicate, consistently, that ProtScan
outperforms DeepBind (wins/ties/loses 20/11/3), that GraphProt and DeepBind
are compatible (W/T/L 11/9/14) and that ProtScan markedly improves on
GraphProt (27/7/0). The average (median) auPRC for the methods was 0.008 (0.006) (ProtScan), 0.003 (0.001) (DeepBind) and 0.002 (0.001) (GraphProt).

\paragraph*{Discussion.}

We can now address the two questions:  Q1) is there an advantage in formulating
the problem  as a regression task rather than a standard classification task?
Q2) is the feature representation obtained by string kernel methods competitive
w.r.t. other state-of-the-art approaches (e.g. based on deep artificial neural
networks)?

A1) The comparison between ProtScan and GraphProt hints at a positive answer.
The two approaches are based on the same technique to map sequences to a vector
space and differ only in the way the task is formulated. We conjecture that the
regression formulation together with the consensus voting and the subsequent
smoothing yields a predictor with a smaller variance and hence a smaller
predictive error. By averaging multiple predictions, we have effectively created
an ensemble classifier where the diversity between predictors is enforced by
training each predictor with sequences at different distances from the target.
Using a single predictor in a regression task, rather than multiple predictors
for each possible distance value, is a way to tie together the various
predictions and can be seen as a form of additional regularization.

A2) Unfortunately we do not have a clear answer to this question. First of all,
the predictive performance of GraphProt and DeepBind (both trained using
only local information around the target regions) is comparable so there is no
clear winner. Secondly, the datasets used in their respective training phases
stem from different experimental methods. While GraphProt models were trained on {\em in vivo} RBP binding sites determined by eCLIP, DeepBind models were trained on RNAcompete data, which consist of short {\em in vitro} RBP-bound sequences.
The effect of a different training regime is therefore a confounder in trying to assess
whether a direct $k$-mer based representation is sufficient or indeed
preferable to the distributed representation employed by neural network
architectures. However, it can stated that the discrete representation is not
markedly worse and is computationally much faster to train.

\subsection{Transcriptome-wide target site modeling}

\begin{figure*}[!tpb]
 \centering
 \includegraphics[width=\textwidth]{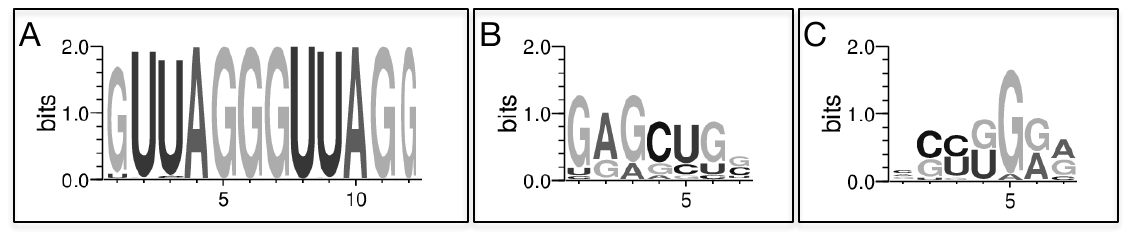}
 \caption{Motifs obtained analyzing the top 5,000 peaks extracted from the
 transcriptome-wide predicted profiles generated by ProtScan. (A) The 12-mer
 GUUAGGGUUAGG resembles the consensus high affinity HNRNPA1 binding site
 UAGGG(A$|$U) identified in~\cite{burd1994rna}. (B-C) The
 7-mer GAGCUGG and all the 6-mers matching the following regular
 expression (C$|$G)(C$|$U)(G$|$U)G(G$|$A)(A$|$G) resemble the consensus motifs
 ACUG and UGGA obtained in~\cite{ascano2012fmrp} analyzing high-throughput data
 of the FMR1 protein.}
 \label{fig:motifs}
\end{figure*}

ProtScan can be used to model RNA-protein interactions and to predict
interaction profiles at a transcriptome-wide scale. As examples we considered the
two vastly studied RBPs HNRNPA1 and FMR1. These RBPs act in different
cellular compartments, i.e. the nucleus for HNRNPA1 and the cytoplasm for FMR1.
Nuclear RBPs, especially splice factors such as HNRNPA1, interact with
pre-(m)RNA that is composed of introns and exons, while cytoplasmic RBPs such
as FMR1 interact with mature RNA molecules, from which the intronic sequences
have been removed during splicing. Note that dealing with mature RNAs and
ignoring intronic sequences shortens the computation time required for
predicting the binding profiles of an order of magnitude.

{\bf HNRNPA1:} HNRNPA1 is member of a family of ubiquitously expressed
heterogeneous nuclear ribonucleoproteins (hnRNPs) which associate with
pre-(m)RNAs in the nucleus and influence their processing, as
well as other aspects of RNA metabolism and transport. HNRNPA1 is one of the
most abundant core proteins of hnRNP complexes and plays a key role in the
regulation of alternative splicing. Mutations in the HNRNPA1 gene have been
observed in individuals with amyotrophic lateral sclerosis
(ALS)~\citep{geuens2016hnrnp}. We considered the eCLIP experiment on K562 cells
(replicate 1) and select HNRNPA1 binding sites with a
fold change threshold of 2.0, resulting in 4,964 interaction sites. We
generated interaction profiles for the entire set of human genes using a
two-fold cross-prediction procedure analogous to the one employed for peak
extraction (using in turn one subset for training and the other for
prediction), obtaining an overall AUC ROC of 0.85.

Next, we extracted the significant peaks from the predicted interaction profiles
as proposed to Section~\ref{sec:peak} and visualized the target regions
identified by ProtScan by running a motif finder procedure on the 5,000 peaks
with the lowest p-value. An {\em in vitro} study by~\cite{burd1994rna}
identified the motif UAGGG(A$|$U) as a consensus high affinity HNRNPA1 binding
site. This consensus sequence is well represented in the HNRNPA1 motif
displayed in Figure~\ref{fig:motifs}A. The 12-mer GUUAGGGUUAGG occurs 63 times
(exact match) in the analyzed subsequences.

{\bf FMR1:} In contrast to HNRNPA1, FMR1 is known to associate with polysomes,
and an expansion of the CGG repeat in the 5' UTR of the FMR1 gene is known to
cause the fragile X syndrome (FXS)~\citep{richter2015dysregulation}. We
considered the eCLIP experiment on K562 cells (replicate 1) and selected
FMR1 binding sites with a fold change threshold of 2.0,
obtaining 26,732 interaction sites. The fact that FMR1 is usually located at
polysomes in the cytoplasm allows us to consider only mature RNAs, i.e. RNAs
without intronic sequences. In humans, alternative splicing enables the
production of more than one transcript from each gene. In order to not
consider every splice variant of each gene, we selected the most prominent
transcript through a series of hierarchical filtering steps: first we considered
the transcript support level (TSL) that identifies well supported transcripts,
then the APPRIS annotation~\citep{rodriguez2012appris} that annotates principal
splicing isoforms, followed by the GENCODE basic annotation that identifies the
representative transcripts of a gene, and finally the transcript length
(preferring longer transcripts). If the procedure ended up producing two or more
transcripts (which are on par on all parameters), the most prominent transcript
was selected at random among them. The selection of the most prominent
transcript for each gene allowed to significantly reduce the size of the
dataset and, therefore, to speed up the prediction of the interaction profiles
for this RBP. Cross-predicted interaction profiles achieved an AUC ROC of 0.79.

As with HNRNPA1, we visualized the FMR1 target regions obtained from the analysis of the
5,000 peaks at lowest p-value. A PAR-CLIP study of FMR1
target sites~\citep{ascano2012fmrp} identified two distinct motifs for
this RBP: ACUG and UGGA. These motifs are in substantial agreement with those
extracted from the ProtScan profiles. The 7-mer GAGCUGG
(Figure~\ref{fig:motifs}B) occured 445 times (exact match) in the considered
subsequences, while the 6-mers matching the following regular expression
(C$|$G)(C$|$U)(G$|$U)G(G$|$A)(A$|$G) (Figure~\ref{fig:motifs}C) were found 4233
times.

The sufficiently high AUC ROC scores indicate that ProtScan can be used to
reliably model the interaction profiles on a transcriptome-wide scale. The
agreement between the resulting motifs and those identified in {\em ad hoc}
studies~\citep{burd1994rna,ascano2012fmrp} additionally supports the quality of the
predicted interaction profiles.

\subsection{Validated predictions} We show that ProtScan can identify valid
candidate locations even in the absence of a significant signal in the CLIP
data for a given region of interest. Here we used the 6 PTBP1 binding sites
predicted by GraphProt on the ANXA7 gene and subsequently verified by mutation
experiments (\cite{ferrarese2014lineage}) (see Supplementary Table 3). We
trained both a GraphProt and a ProtScan model on the available  PTBP1 K562
eCLIP binding sites. For model training, 2718 positive sites were used for
GraphProt (fold change > 4.2) and 11630 for ProtScan (fold change > 3). Note
however that the PTBP1 eCLIP quality does not satisfy the requirements used in
the previous experiments, since the binding site are not sufficiently
consistent across replicates. Notwithstanding the lower quality,  ProtScan can
learn a reliable predictive model. In more details, nucleotide wise binding
scores on the full ANXA7 gene sequence (taken from Ensembl human genes
GRCh38.p10) were predicted for both methods. Nucleotide wise predictions were
additionally calculated on a set of 5000 randomly selected transcript
sequences (same Ensembl release) in order to generate an empirical
distribution for the background model. To extract top-scoring regions, we only
considered contiguous regions with p-value < 0.05 and checked the overlap with
the experimentally verified PTBP1 binding sites. For GraphProt, 9 of a set of
368 regions deemed significant overlapped with 6 of 6 verified PTBP1 binding
sites, while for ProtScan 3 of 34 proposed regions were overlapping with 3 of
6 verified PTBP1 binding sites. These results indicate that while GraphProt's
recall is higher it also tends to predict a larger number of locations than
ProtScan and is therefore likely to exhibit a lower precision. Equating the
non verified proposals sites to false positives we can compute a balanced
F-measure (i.e. the harmonic mean of precision and recall) of 0.05 for
GraphProt and three times larger (0.15) for ProtScan.

\section{Conclusion}

In this work we presented ProtScan, a tool based on kernelized regression and
consensus voting for modeling and predicting RNA-protein interaction profiles.
Extensive experimental results show that the tool is competitive w.r.t.
two state-of-the-art
approaches~\citep{maticzka2014graphprot,alipanahi2015predicting}, and that it
can reliably be used in a transcriptome-wide setting. In addition to the
software, we release 146 ProtScan models for 46 RBPs, induced from eCLIP
experiments performed in different cell lines (HepG2 and K562), in dual
replicates, and considering different fold change thresholds for selecting the
binding sites for training (see Section~\ref{sec:dataset}).

To further improve the performance of ProtScan, future implementations could
include additional types of information, which are known to be associated with
RBP binding, including mRNA accessibility and the presence of target sites for
regulatory entities such as miRNAs and other known competitive or cooperative
RBPs.



\section*{Funding}

M.U. and R.B. are funded by the Baden-Wuerttemberg-Stiftung (BWST\_NCRNA\_008)
and the German Research Foundation (DFG grant BA2168/11-1 SPP
1738).\vspace*{-12pt}

\bibliographystyle{natbib}

\begin{thebibliography}{}

\bibitem[Alipanahi {\em et~al.}(2015)Alipanahi {\em
  et~al.}]{alipanahi2015predicting}
Alipanahi, B. {\em et~al.} (2015).
\newblock {Predicting the sequence specificities of DNA-and RNA-binding
  proteins by deep learning}.
\newblock {\em Nature biotechnology\/}, {\bf 33}(8), 831--838.

\bibitem[Anders {\em et~al.}(2011)Anders, Mackowiak, Jens, Maaskola, Kuntzagk,
  Rajewsky, Landthaler, and Dieterich]{anders2011dorina}
Anders, G., Mackowiak, S.~D., Jens, M., Maaskola, J., Kuntzagk, A., Rajewsky,
  N., Landthaler, M., and Dieterich, C. (2011).
\newblock dorina: a database of rna interactions in post-transcriptional
  regulation.
\newblock {\em Nucleic acids research\/}, {\bf 40}(D1), D180--D186.

\bibitem[Ascano {\em et~al.}(2012)Ascano {\em et~al.}]{ascano2012fmrp}
Ascano, M. {\em et~al.} (2012).
\newblock {FMRP targets distinct mRNA sequence elements to regulate protein
  expression}.
\newblock {\em Nature\/}, {\bf 492}(7429), 382--386.

\bibitem[Bailey(2011)Bailey]{bailey2011dreme}
Bailey, T.~L. (2011).
\newblock Dreme: motif discovery in transcription factor chip-seq data.
\newblock {\em Bioinformatics\/}, {\bf 27}(12), 1653--1659.

\bibitem[Bailey {\em et~al.}(1994)Bailey, Elkan, {\em
  et~al.}]{bailey1994fitting}
Bailey, T.~L., Elkan, C., {\em et~al.} (1994).
\newblock Fitting a mixture model by expectation maximization to discover
  motifs in bipolymers.

\bibitem[Baltz {\em et~al.}(2012)Baltz {\em et~al.}]{baltz2012mrna}
Baltz, A.~G. {\em et~al.} (2012).
\newblock {The mRNA-bound proteome and its global occupancy profile on
  protein-coding transcripts}.
\newblock {\em Molecular cell\/}, {\bf 46}(5), 674--690.

\bibitem[Bergstra and Bengio(2012)Bergstra and Bengio]{bergstra2012random}
Bergstra, J. and Bengio, Y. (2012).
\newblock {Random search for hyper-parameter optimization}.
\newblock {\em Journal of Machine Learning Research\/}, {\bf 13}(Feb),
  281--305.

\bibitem[Breiman(2001)Breiman]{breiman2001random}
Breiman, L. (2001).
\newblock {Random forests}.
\newblock {\em Machine learning\/}, {\bf 45}(1), 5--32.

\bibitem[Burd and Dreyfuss(1994)Burd and Dreyfuss]{burd1994rna}
Burd, C.~G. and Dreyfuss, G. (1994).
\newblock {RNA binding specificity of hnRNP A1: significance of hnRNP A1
  high-affinity binding sites in pre-mRNA splicing.}
\newblock {\em The EMBO journal\/}, {\bf 13}(5), 1197.

\bibitem[Castello {\em et~al.}(2012)Castello {\em
  et~al.}]{castello2012insights}
Castello, A. {\em et~al.} (2012).
\newblock {Insights into RNA biology from an atlas of mammalian mRNA-binding
  proteins}.
\newblock {\em Cell\/}, {\bf 149}(6), 1393--1406.

\bibitem[Comaniciu and Meer(2002)Comaniciu and Meer]{comaniciu2002mean}
Comaniciu, D. and Meer, P. (2002).
\newblock {Mean shift: A robust approach toward feature space analysis}.
\newblock {\em IEEE Transactions on pattern analysis and machine
  intelligence\/}, {\bf 24}(5), 603--619.

\bibitem[Cook {\em et~al.}(2010)Cook, Kazan, Zuberi, Morris, and
  Hughes]{cook2010rbpdb}
Cook, K.~B., Kazan, H., Zuberi, K., Morris, Q., and Hughes, T.~R. (2010).
\newblock Rbpdb: a database of rna-binding specificities.
\newblock {\em Nucleic acids research\/}, {\bf 39}(suppl\_1), D301--D308.

\bibitem[Corcoran {\em et~al.}(2011)Corcoran {\em
  et~al.}]{corcoran2011paralyzer}
Corcoran, D.~L. {\em et~al.} (2011).
\newblock {PARalyzer: definition of RNA binding sites from PAR-CLIP short-read
  sequence data}.
\newblock {\em Genome biology\/}, {\bf 12}(8), R79.

\bibitem[Costa and De~Grave(2010)Costa and De~Grave]{costa2010fast}
Costa, F. and De~Grave, K. (2010).
\newblock {Fast neighborhood subgraph pairwise distance kernel}.
\newblock In {\em Proceedings of the 26th International Conference on Machine
  Learning\/}, pages 255--262. Omnipress.

\bibitem[Das and Dai(2007)Das and Dai]{das2007survey}
Das, M.~K. and Dai, H.-K. (2007).
\newblock A survey of dna motif finding algorithms.
\newblock {\em BMC bioinformatics\/}, {\bf 8}(7), S21.

\bibitem[Ferrarese {\em et~al.}(2014)Ferrarese {\em
  et~al.}]{ferrarese2014lineage}
Ferrarese, R. {\em et~al.} (2014).
\newblock {Lineage-specific splicing of a brain-enriched alternative exon
  promotes glioblastoma progression}.
\newblock {\em The Journal of clinical investigation\/}, {\bf 124}(7),
  2861--2876.

\bibitem[Foat {\em et~al.}(2006)Foat, Morozov, and
  Bussemaker]{foat2006statistical}
Foat, B.~C., Morozov, A.~V., and Bussemaker, H.~J. (2006).
\newblock Statistical mechanical modeling of genome-wide transcription factor
  occupancy data by matrixreduce.
\newblock {\em Bioinformatics\/}, {\bf 22}(14), e141--e149.

\bibitem[Gerstberger {\em et~al.}(2014)Gerstberger {\em
  et~al.}]{gerstberger2014census}
Gerstberger, S. {\em et~al.} (2014).
\newblock {A census of human RNA-binding proteins}.
\newblock {\em Nature Reviews Genetics\/}, {\bf 15}(12), 829--845.

\bibitem[Geuens {\em et~al.}(2016)Geuens {\em et~al.}]{geuens2016hnrnp}
Geuens, T. {\em et~al.} (2016).
\newblock {The hnRNP family: insights into their role in health and disease}.
\newblock {\em Human genetics\/}, pages 1--17.

\bibitem[Gupta {\em et~al.}(2013)Gupta, Kosti, Plaut, Pivko, Tkacz,
  Cohen-Chalamish, Biswas, Wachtel, Waldman Ben-Asher, Carmi, {\em
  et~al.}]{gupta2013hnrnp}
Gupta, S.~K., Kosti, I., Plaut, G., Pivko, A., Tkacz, I.~D., Cohen-Chalamish,
  S., Biswas, D.~K., Wachtel, C., Waldman Ben-Asher, H., Carmi, S., {\em
  et~al.} (2013).
\newblock The hnrnp f/h homologue of trypanosoma brucei is differentially
  expressed in the two life cycle stages of the parasite and regulates splicing
  and mrna stability.
\newblock {\em Nucleic acids research\/}, {\bf 41}(13), 6577--6594.

\bibitem[Hafner {\em et~al.}(2010)Hafner {\em et~al.}]{hafner2010transcriptome}
Hafner, M. {\em et~al.} (2010).
\newblock {Transcriptome-wide identification of RNA-binding protein and
  microRNA target sites by PAR-CLIP}.
\newblock {\em Cell\/}, {\bf 141}(1), 129--141.

\bibitem[Hansen and Salamon(1990)Hansen and Salamon]{hansen1990neural}
Hansen, L.~K. and Salamon, P. (1990).
\newblock {Neural network ensembles}.
\newblock {\em IEEE transactions on pattern analysis and machine
  intelligence\/}, {\bf 12}(10), 993--1001.

\bibitem[Hiller {\em et~al.}(2006)Hiller, Pudimat, Busch, and
  Backofen]{hiller2006using}
Hiller, M., Pudimat, R., Busch, A., and Backofen, R. (2006).
\newblock Using rna secondary structures to guide sequence motif finding
  towards single-stranded regions.
\newblock {\em Nucleic acids research\/}, {\bf 34}(17), e117--e117.

\bibitem[Kazan and Morris(2013)Kazan and Morris]{kazan2013rbpmotif}
Kazan, H. and Morris, Q. (2013).
\newblock Rbpmotif: a web server for the discovery of sequence and structure
  preferences of rna-binding proteins.
\newblock {\em Nucleic acids research\/}, {\bf 41}(W1), W180--W186.

\bibitem[Kazan {\em et~al.}(2010)Kazan, Ray, Chan, Hughes, and
  Morris]{kazan2010rnacontext}
Kazan, H., Ray, D., Chan, E.~T., Hughes, T.~R., and Morris, Q. (2010).
\newblock Rnacontext: a new method for learning the sequence and structure
  binding preferences of rna-binding proteins.
\newblock {\em PLoS computational biology\/}, {\bf 6}(7), e1000832.

\bibitem[Khorshid {\em et~al.}(2010)Khorshid, Rodak, and
  Zavolan]{khorshid2010clipz}
Khorshid, M., Rodak, C., and Zavolan, M. (2010).
\newblock Clipz: a database and analysis environment for experimentally
  determined binding sites of rna-binding proteins.
\newblock {\em Nucleic acids research\/}, {\bf 39}(suppl\_1), D245--D252.

\bibitem[K{\"o}nig {\em et~al.}(2010)K{\"o}nig {\em et~al.}]{konig2010iclip}
K{\"o}nig, J. {\em et~al.} (2010).
\newblock {iCLIP reveals the function of hnRNP particles in splicing at
  individual nucleotide resolution}.
\newblock {\em Nature structural \& molecular biology\/}, {\bf 17}(7),
  909--915.

\bibitem[Krogh {\em et~al.}(1995)Krogh {\em et~al.}]{krogh1995neural}
Krogh, A. {\em et~al.} (1995).
\newblock {Neural network ensembles, cross validation, and active learning}.
\newblock {\em Advances in neural information processing systems\/}, {\bf 7},
  231--238.

\bibitem[Leibovich {\em et~al.}(2013)Leibovich, Paz, Yakhini, and
  Mandel-Gutfreund]{leibovich2013drimust}
Leibovich, L., Paz, I., Yakhini, Z., and Mandel-Gutfreund, Y. (2013).
\newblock Drimust: a web server for discovering rank imbalanced motifs using
  suffix trees.
\newblock {\em Nucleic acids research\/}, {\bf 41}(W1), W174--W179.

\bibitem[Leslie {\em et~al.}(2002)Leslie {\em et~al.}]{leslie2002spectrum}
Leslie, C.~S. {\em et~al.} (2002).
\newblock {The spectrum kernel: A string kernel for SVM protein
  classification.}
\newblock In {\em Pacific symposium on biocomputing\/}, volume~7, pages
  566--575.

\bibitem[Licatalosi {\em et~al.}(2008)Licatalosi {\em
  et~al.}]{licatalosi2008hits}
Licatalosi, D.~D. {\em et~al.} (2008).
\newblock {HITS-CLIP yields genome-wide insights into brain alternative RNA
  processing}.
\newblock {\em Nature\/}, {\bf 456}(7221), 464--469.

\bibitem[Lovci {\em et~al.}(2013)Lovci {\em et~al.}]{lovci2013rbfox}
Lovci, M.~T. {\em et~al.} (2013).
\newblock {Rbfox proteins regulate alternative mRNA splicing through
  evolutionarily conserved RNA bridges}.
\newblock {\em Nature structural \& molecular biology\/}, {\bf 20}(12),
  1434--1442.

\bibitem[Maticzka {\em et~al.}(2014)Maticzka {\em
  et~al.}]{maticzka2014graphprot}
Maticzka, D. {\em et~al.} (2014).
\newblock {GraphProt: modeling binding preferences of RNA-binding proteins}.
\newblock {\em Genome biology\/}, {\bf 15}(1), R17.

\bibitem[Patel {\em et~al.}(2012)Patel {\em et~al.}]{patel2012spatial}
Patel, V.~L. {\em et~al.} (2012).
\newblock {Spatial arrangement of an RNA zipcode identifies mRNAs under
  post-transcriptional control}.
\newblock {\em Genes \& Development\/}, {\bf 26}(1), 43--53.

\bibitem[Perrone(1993)Perrone]{perrone1993improving}
Perrone, M.~P. (1993).
\newblock {\em {Improving regression estimation: Averaging methods for variance
  reduction with extensions to general convex measure optimization}\/}.
\newblock Ph.D. thesis, Brown University.

\bibitem[Ray {\em et~al.}(2009)Ray {\em et~al.}]{ray2009rapid}
Ray, D. {\em et~al.} (2009).
\newblock {Rapid and systematic analysis of the RNA recognition specificities
  of {RNA}-binding proteins}.
\newblock {\em Nature biotechnology\/}, {\bf 27}(7), 667--670.

\bibitem[Ray {\em et~al.}(2013)Ray {\em et~al.}]{ray2013compendium}
Ray, D. {\em et~al.} (2013).
\newblock {A compendium of RNA-binding motifs for decoding gene regulation}.
\newblock {\em Nature\/}, {\bf 499}(7457), 172--177.

\bibitem[Richter {\em et~al.}(2015)Richter {\em
  et~al.}]{richter2015dysregulation}
Richter, J.~D. {\em et~al.} (2015).
\newblock {Dysregulation and restoration of translational homeostasis in
  fragile X syndrome}.
\newblock {\em Nature Reviews Neuroscience\/}.

\bibitem[Rodriguez {\em et~al.}(2012)Rodriguez {\em
  et~al.}]{rodriguez2012appris}
Rodriguez, J.~M. {\em et~al.} (2012).
\newblock {APPRIS: annotation of principal and alternative splice isoforms}.
\newblock {\em Nucleic acids research\/}, page gks1058.

\bibitem[Sanford {\em et~al.}(2009)Sanford, Wang, Mort, VanDuyn, Cooper,
  Mooney, Edenberg, and Liu]{sanford2009splicing}
Sanford, J.~R., Wang, X., Mort, M., VanDuyn, N., Cooper, D.~N., Mooney, S.~D.,
  Edenberg, H.~J., and Liu, Y. (2009).
\newblock Splicing factor sfrs1 recognizes a functionally diverse landscape of
  rna transcripts.
\newblock {\em Genome research\/}, {\bf 19}(3), 381--394.

\bibitem[Shawe-Taylor and Cristianini(2004)Shawe-Taylor and
  Cristianini]{shawe2004kernel}
Shawe-Taylor, J. and Cristianini, N. (2004).
\newblock {\em {Kernel methods for pattern analysis}\/}.
\newblock Cambridge university press.

\bibitem[Sloan {\em et~al.}(2016)Sloan {\em et~al.}]{sloan2016encode}
Sloan, C.~A. {\em et~al.} (2016).
\newblock {ENCODE data at the ENCODE portal}.
\newblock {\em Nucleic acids research\/}, {\bf 44}(D1), D726--D732.

\bibitem[Speir {\em et~al.}(2016)Speir {\em et~al.}]{speir2016ucsc}
Speir, M.~L. {\em et~al.} (2016).
\newblock {The UCSC genome browser database: 2016 update}.
\newblock {\em Nucleic acids research\/}, {\bf 44}(D1), D717--D725.

\bibitem[Uren {\em et~al.}(2012)Uren {\em et~al.}]{uren2012site}
Uren, P.~J. {\em et~al.} (2012).
\newblock {Site identification in high-throughput RNA--protein interaction
  data}.
\newblock {\em Bioinformatics\/}, {\bf 28}(23), 3013--3020.

\bibitem[Van~Nostrand {\em et~al.}(2016)Van~Nostrand {\em
  et~al.}]{van2016robust}
Van~Nostrand, E.~L. {\em et~al.} (2016).
\newblock {Robust transcriptome-wide discovery of RNA-binding protein binding
  sites with enhanced CLIP (eCLIP)}.
\newblock {\em Nature methods\/}, {\bf 13}(6), 508--514.

\bibitem[Vaquerizas {\em et~al.}(2009)Vaquerizas {\em
  et~al.}]{vaquerizas2009census}
Vaquerizas, J.~M. {\em et~al.} (2009).
\newblock {A census of human transcription factors: function, expression and
  evolution}.
\newblock {\em Nature Reviews Genetics\/}, {\bf 10}(4), 252--263.

\bibitem[Yates {\em et~al.}(2015)Yates {\em et~al.}]{yates2015ensembl}
Yates, A. {\em et~al.} (2015).
\newblock {Ensembl 2016}.
\newblock {\em Nucleic acids research\/}, page gkv1157.

\bibitem[Zhang {\em et~al.}(2013)Zhang, Lee, Swanson, and
  Darnell]{zhang2013prediction}
Zhang, C., Lee, K.-Y., Swanson, M.~S., and Darnell, R.~B. (2013).
\newblock Prediction of clustered rna-binding protein motif sites in the
  mammalian genome.
\newblock {\em Nucleic acids research\/}, {\bf 41}(14), 6793--6807.

\end{thebibliography}

\end{document}